\documentclass[twocolumn,pre,floatfix]{revtex4}
\usepackage{graphicx}

\newcommand{\kB}{k_{\mathrm{B}}}
\newcommand{\kT}{\kB T}
\newcommand{\excess}{{\mathrm{(ex)}}}

\newcommand{\fnex}{f_N^\excess}
\newcommand{\fnexp}{\fnex{}'}

\newcommand{\rc}{r_{{c}}}
\newcommand{\rd}{r_{{d}}}
\newcommand{\rvec}{{\mathbf{r}}}
\newcommand{\vvec}{{\mathbf{v}}}
\newcommand{\vij}{\vvec_{ij}}
\newcommand{\Fvec}{{\mathbf{F}}}
\newcommand{\Fiext}{\Fvec_{i,{\mathrm{ext}}}}
\newcommand{\Fij}{\Fvec_{ij}}
\newcommand{\FijC}{\Fij^{\mathrm{C}}}
\newcommand{\FijD}{\Fij^{\mathrm{D}}}
\newcommand{\FijR}{\Fij^{\mathrm{R}}}
\newcommand{\wC}{w_{\mathrm{C}}}
\newcommand{\wD}{w_{\mathrm{D}}}
\newcommand{\wR}{w_{\mathrm{R}}}
\newcommand{\evec}{{\mathbf{e}}}
\newcommand{\eij}{\evec_{ij}}
\newcommand{\eqref}[1]{(\ref{#1})}
\newcommand{\wrho}{w_{\rho}}
\newcommand{\rhobar}{\overline{\rho}}
\newcommand{\rhobarbar}{{\overline{\overline{\rho}}}}
\newcommand{\rhoL}{\rho_{\mathrm{L}}}
\newcommand{\rhoV}{\rho_{\mathrm{V}}}
\newcommand{\alphaMF}{\alpha_{\mathrm{MF}}}
\newcommand{\pMF}{p_{\mathrm{MF}}}
\newcommand{\average}[1]{\langle #1\rangle}

\newcommand{\etal}{\emph{et al}}

\begin{document}

\title{Vapour-liquid coexistence in many-body dissipative particle dynamics}

\author{P. B. Warren}
\affiliation{Unilever Research and Development Port Sunlight, 
Bebington, Wirral, CH63 3JW, UK.}

\date{May 20, 2003}

\begin{abstract}
Many-body dissipative particle dynamics is constructed to exhibit
vapour-liquid coexistence, with a sharp interface, and a vapour phase
of vanishingly small density.  In this form, the model is an unusual
example of a soft-sphere liquid with a potential energy built out of
local-density dependent one-particle self energies.  The application to
fluid mechanics problems involving free surfaces is illustrated by
simulation of a pendant drop.
\end{abstract}

\maketitle

\section{Introduction}
Dissipative particle dynamics (DPD) is familiar as a method of
simulating complex fluids at a coarse grained level \cite{HK,GW}, for
example block copolymer polymer melts \cite{GM,GMT}, and surfactant
solutions \cite{JBCKHRW,PWM}.  DPD has also been used for multiphase
fluid problems, such as phase separation kinetics in binary liquid
mixtures \cite{CN,JBKC,NC}, droplet deformation and rupture in shear
fields \cite{CLRW}, and droplets on surfaces under the influence of
shear fields \cite{JLRS}.  The advantage of DPD for these kind of
problems lies in the simplicity of the underlying algorithm, and the
physical way in which singular events such as droplet rupture are
captured.  Such considerations also make the method attractive for
\emph{free-surface} fluid dynamics problems.  Examples of these
include various kinds of wetting, spreading, wicking, and capillary
problems.  To be used for these kind of problems, DPD needs to be
extended to allow for vapour-liquid equilibrium.  In this way a free
surface will arise naturally as a vapour-liquid interface, and such an
interface will possess the same physics as a clean vapour-liquid
interface.

To achieve vapour-liquid coexistence in DPD, for a single component
system, requires a van der Waals loop in the equation of state (EOS)
(pressure-density curve).  However this presents a fundamental
limitation for standard DPD since the soft interaction forces used in
the method invariably lead to a predominantly quadratic EOS \cite{LBH}.
One way around this is the `many-body' DPD method invented by
Pagonabarraga and Frenkel \cite{PF,RDGnote}, and also investigated by
Trofimov \etal\ \cite{TNM}.  In many-body DPD, the amplitude of the soft
repulsions is made to depend on the local density.  In this way one
can achieve a much wider range of possibilities for the EOS.

In the present work, many-body DPD is developed to exhibit
vapour-liquid coexistence, with a sharp interface, and a vapour phase
of vanishingly small density.  The approach taken is fundamentally the
same approach as used in Refs.~\onlinecite{PF} and \onlinecite{TNM},
but with a somewhat different interpretation of the same mathematics.
Therefore a general theory for many-body DPD is described first, which
I argue involves a fundamental re-interpretation of the DPD
interaction potentials.  The specific implementation for vapour-liquid
equilibrium is described next, and finally the application to free
surface problems is illustrated by simulation of a pendant droplet.
Another application to vapour-liquid phase separation kinetics was
described in an earlier note \cite{WarrenPRL}.

\section{General theory}
Dissipative particle dynamics (DPD) is basically Molecular
Dynamics \cite{ATbook,FSbook}, with two key innovations.  The first,
and perhaps the most profound, is the use of soft interactions.  This
stands in contrast to the common use of interaction potentials
corresponding to particles with hard cores, for example Lennard-Jones
interactions or modified hard-sphere interactions.  The second
innovation is the use of a momentum conserving thermostat.  This
allows one to simulate at a well defined temperature yet preserve
hydrodynamics, and this can be important for some problems such as
phase separation kinetics.  The thermostat described below is the
original (Espa\~nol-Warren) thermostat \cite{EW}, although the
Lowe-Anderson thermostat is perhaps simpler and more
efficient \cite{Lowe}.  In the present paper, the focus is on the
equilibrium properties of many-body DPD models, and the nature of the
thermostat is unimportant.

The particles in DPD have positions $\rvec_i$ and velocities
$\vvec_i$, where $i = 1$ to $N$ runs over the set of particles, moving
in a simulation box of volume $V$.  They move according to the
kinematic condition $d\rvec_i/dt = \vvec_i$, and Newton's second law
$d\vvec_i/dt = \Fvec_i/m_i$ where $m_i$ is the mass of the $i$th
particle.  Here
\begin{equation}
\Fvec_i = \Fiext + {\textstyle \sum_{j\ne i}} \,\Fij
\end{equation}
is the total force acting on the $i$th particle, comprising a possible
external force $\Fiext$ and forces $\Fij$ due to the interaction
between the $i$th and $j$th particles.  The interaction forces are
decomposed into conservative, dissipative and random contributions,
\begin{equation}
\Fij = \FijC + \FijD + \FijR.  
\end{equation}
The individual contributions all vanish for particle separations
larger than some cutoff interaction range $\rc$, and all obey
Newton's third law so that $\Fij + \Fvec_{ji} = 0$.

The conservative force is 
\begin{equation}
\FijC = A\,\wC(r_{ij})\,\eij
\label{feq}
\end{equation}
where $\rvec_{ij} = \rvec_j - \rvec_i$, $r_{ij} = |\rvec_{ij}|$, and
$\eij = \rvec_{ij}/r_{ij}$.  The weight function $\wC(r)$ vanishes for
$r > \rc$, and for simplicity is taken to decreases linearly with
particle separation, thus $\wC(r) = (1-r/\rc)$.  This force
corresponds to a total potential energy which is a sum of pair
potentials
\begin{equation}
U(\lbrace\rvec_i\rbrace) = {\textstyle \sum_{j>i}} \,\phi(r_{ij})
\label{ueq}
\end{equation}
where $-\phi\,{}'(r) = A \wC(r)$, and thus $\phi(r)=(A/2)(1-r/\rc)^2$
for standard DPD.  The dissipative and random forces are $\FijD =
-\gamma \wD(r_{ij}) (\vij \cdot \eij ) \eij$ and $\FijR = \sigma
\wR(r_{ij}) \xi_{ij} \eij$.  In these $\gamma$ and $\sigma$ are
amplitudes, $\wD(r)$ and $\wR(r)$ are additional weight functions also
vanishing for $r > \rc$, $\vij = \vvec_j - \vvec_i$, and $\xi_{ij} =
\xi_{ji}$ is pairwise continuous white noise with
$\langle\xi_{ij}(t)\rangle = 0$ and
$\langle\xi_{ij}(t)\xi_{kl}(t')\rangle = (\delta_{ik}\delta_{jl} +
\delta_{il}\delta_{jk})\delta(t-t')$.  The dissipative and random
forces act as the above-mentioned thermostat provided that the weight
functions and amplitudes are chosen to obey a fluctuation-dissipation
theorem: $\sigma^2 = 2\gamma\kT$ and $\wD = (\wR)^2$, where $\kT$ is
the desired temperature in units of Boltzmann's constant
$\kB$ \cite{EW}. The same weight function is used as for the
conservative forces (basically for historical reasons): $\wR = \wC$
and $\wD = (\wC)^2$.

Usually all the particles are assumed to have the same mass, and to
fix units of mass and length a convenient choice is to set $m_i = \rc
= 1$.  Often the units of energy and hence time are fixed by setting
$\kT = 1$, but for equilibrium simulations it can be convenient to
keep $\kT$ as a free parameter.

The integration of the equations of motion is a non-trivial matter
since one has to manage the random forces somehow.  For an integration
algorithm, Groot and Warren investigated a version of the
velocity-Verlet scheme used in molecular dynamics
simulations \cite{GW}, but it was later shown by den Otter and Clarke
that this is not a real improvement over a simple Euler type
integration scheme \cite{dOC}. More extensive studies have been
undertaken by Vattulainen \etal\ \cite{VKBP}.  But many the problems are
obviated if the Lowe-Anderson thermostat is used, which is based on
distinctly different physical ideas \cite{Lowe}.  All the simulations
described below though were carried out with the simple
velocity-Verlet like algorithm described by Groot and Warren.

For a single-component DPD fluid, the equation of state (EOS) gives
the pressure $p$ as a function of the density $\rho = N/V$.  For the
soft potential given above, the EOS is now well established to
be \cite{GW}
\begin{equation}
p = \rho\kT + \alpha A\rho^2
\label{eqpstd}
\end{equation}
where $\alpha = 0.101 \pm 0.001$ is very close to the mean field
prediction $\alphaMF = \pi/30 = 0.1047$ (see also below). The first
term in the EOS is an ideal gas term, and the second term is the
excess pressure, which is almost perfectly quadratic in the density
(there is a very small correction of order $\rho^3$).  Note though
that $\alpha A$ is \emph{not} the second virial coefficient \cite{LBH},
so the above EOS breaks down as $\rho\to0$.  It seems that a quadratic
EOS like this is unavoidable for soft potentials \cite{LBH}.  This
represents the fundamental limitation to basic DPD mentioned in the
introduction.  Moreover one has to take $A \ge 0$ otherwise the
pressure diverges negatively at high densities, so one is are
restricted to a strictly positive compressibility $\partial
p/\partial\rho > 0$.  In fact, making $A < 0$ throws the DPD pair
potential into a formal class of catastrophic potentials for which it
can be rigorously proved there is no thermodynamic
limit \cite{LBH,RuelleBook}.  The situation is not as grim as it might
seem though since considerable progress can be made for applications
by introducing different species of particles, and allowing them to be
differentiated by their repulsion amplitudes thus $A\to A_{ij}$ in
Eqs.~\eqref{feq} and \eqref{ueq}.

For the one-component fluid, an obvious way to get around the problem
of a quadratic EOS is to make the amplitude $A$ in the force law
dependent on density somehow.  Such a scheme has been examined by
several workers \cite{PF,TNM}, and proves to be a simple extension to
DPD.  This many-body DPD requires only a modest additional
computational cost, but throws open the possibility to simulate
systems with an arbitrarily complicated EOS.  Care must be taken
though with density-dependent interactions \cite{Louis}.  The approach
described here introduces a local density into the amplitude in the
force law.  By being explicit about the construction of the local
density, this is a `safe' way to introduce a density-dependence into
the interations \cite{ALRT}.

In many-body DPD, I write
\begin{equation}
\FijC = \frac{1}{2}[A(\rhobar_i) + A(\rhobar_j)]\,\wC(r_{ij})\,\eij, 
\label{mbfeq}
\end{equation}
for a one-component fluid (Trofimov \etal\ describe a
multi-component generalisation).  A partial amplitude $A(\rhobar)$ is
introduced, depending on a weighted local density, which I define for
the $i$th particle to be
\begin{equation}
\rhobar_i = {\textstyle\sum_{j\ne i}} \,\wrho(r_{ij}).
\label{rhobareq}
\end{equation}
The weight function $\wrho(r)$ vanishes for $r > \rc$ and for
convenience is normalised so that $\int d^3\rvec\, \wrho(r) = 1$,
although in principle the normalisation could be absorbed into the
definition of $A(\rhobar_i)$.  The discounted self contribution $i=j$
in Eq.~\eqref{rhobareq} would only add a constant $\wrho(0)$ to
$\rhobar_i$, amounting to a constant shift of the argument in the
definition of $A(\rhobar_i)$ (see Trofimov \etal\ for a more extensive
discussion on this point).  The weighted local density is readily
computed by an additional sweep through the neighbour list, hence
there is only a modest additional computational overhead.  If
$A(\rhobar) = A$, the method reduces exactly to the standard DPD
model.  

In mean field theory, it is easy to show that the modified force law
should give an EOS
\begin{equation}
\pMF = \rho\kT + \alphaMF A(\rho)\rho^2
\label{mbpeq}
\end{equation}
where 
\begin{equation}
\alphaMF=({2\pi}/{3}){\textstyle\int_0^\infty} dr\,r^3\,\wC(r),
\label{amfeq}
\end{equation}
(ie $\alphaMF=\pi/30$ for the standard choice of $\wC(r)$).  Thus, in
principle, an arbitrary dependence on density can be recovered.

This is not the end of the story though.  The existence of a potential
energy $U(\lbrace\rvec_i\rbrace)$ such that $\Fvec_i = \partial
U/\partial\rvec_i$ requires the forces to obey a `Maxwell relation' of
the type $\partial\Fvec_i/\partial\rvec_j =
\partial\Fvec_j/\partial\rvec_i$.  This is a non-trivial requirement
since the particle positions appear both directly in Eq.~\eqref{mbfeq}
and indirectly through the definition of the local density in
Eq.~\eqref{rhobareq}.  One can show a necessary and sufficient
condition for it to be true is that the derivative $\wrho'$ is
proportional to $\wC$, so the two weight functions $\wrho$ and $\wC$
are not independent.  One can then prove from the normalisation
condition on $\wrho(r)$ that
\begin{equation}
-\wrho'(r)=\wC(r) / 2\alphaMF
\label{maxeq}
\end{equation}
where $\alphaMF$ is defined in Eq.~\eqref{amfeq}.

What, then, is the corresponding potential?  I find
that 
\begin{equation}
U(\lbrace\rvec_i\rbrace) = {\textstyle \sum_i} \,u(\rhobar_i)
\label{mbueq}
\end{equation}
where $u(\rhobar_i)$ is a self energy depending on the local density,
such that 
\begin{equation}
u'(\rhobar) = \alphaMF A(\rhobar).
\end{equation}
Comparing Eq.~\eqref{mbueq} with Eq.~\eqref{ueq}, it is clear that
there has been a \emph{profound shift in perspective}, from a
potential function expressed in terms of soft pair potentials, to one
expressed in terms of density-dependent self energies.

There have recently been many discussions on the thermodynamic
consistency of density-dependent interactions in the
literature \cite{Louis,SST,Tejero,TB}.  However, for the present
formulation all thermodynamic relations are valid because the
underlying potential $U(\lbrace\rvec_i\rbrace)$ is a well defined,
density-independent function of the particle positions.  This is
important because it means for instance the virial equation for the
pressure or stress tensor, constructed out of the forces, can be used
without change.

If $u(\rhobar)$ is a polynomial in $\rhobar$ of order $n$, it is easy
to show that $\sum_i u(\rhobar_i)$ expands to a sum over
$(n+1)$-body density-independent potentials.  For $u(\rhobar) =
\alphaMF A\rhobar$, standard DPD is recovered.

I wish to emphasise, contrary to some hints in the
literature \cite{PF,TNM}, that $\average{U(\lbrace\rvec_i\rbrace)}$ is
the internal energy and \emph{not} the excess free energy (here
$\average{\dots}$ is a thermal average).  It follows from
Eqs.~\eqref{mbpeq} and \eqref{mbueq} that the mean field EOS is $\pMF
= \rho\kT + \rho^2 u'(\rho)$.  A standard thermodynamic result for the
true pressure is $p = \rho\kT + \rho^2 \fnexp(\rho)$ where
$\fnex(\rho)$ is the excess free energy per particle.  This shows that
the interpretation $u\equiv\fnex$ is a mean field approximation, and
as such will be spoilt by correlation effects.  Note also that
correlations mean, typically, $\average{\rhobar_i} \ne \rho$, and
$\average{u(\rhobar_i)} \ne u(\average{\rhobar_i}) \ne u(\rho)$.
Trofimov \etal\ give results for the mean local density, and suggest
ways that one might improve the correspondence between
$\average{\rhobar_i}$ and $\rho$.  Here I take a different approach,
and regard $\rhobar_i$ as a convenient intermediate quantity which is
used to construct the forces; as such it is not important that its
average differs from $\rho$.  In practice, like Trofimov \etal, I find
that the mean field EOS for many-body DPD can be considerably less
accurate compared to standard DPD.  Thus the method always requires
calibration to determine the true thermodynamic properties (the
approach of Trofimov \etal\ can be used to achieve a specific EOS).

\begin{figure}
\begin{center}
(a)\includegraphics{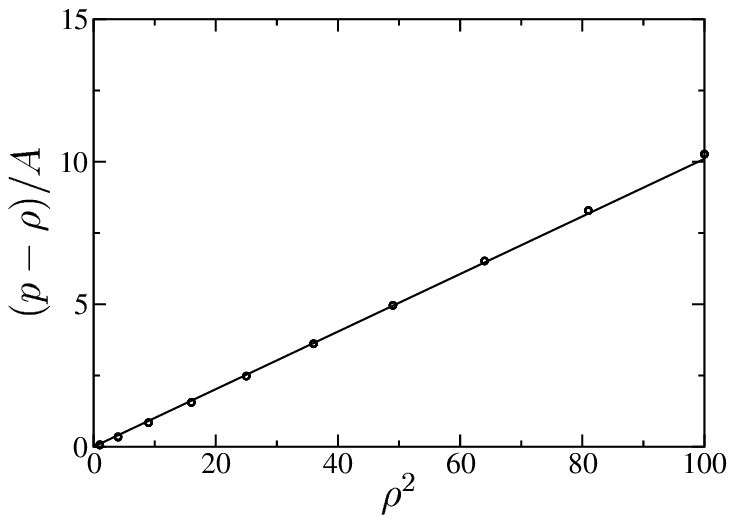}\\
(b)\includegraphics{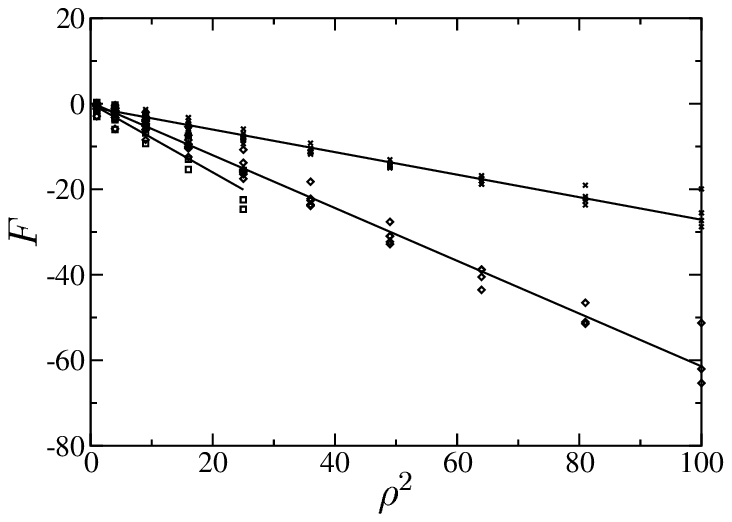}
\end{center}
\caption[?]{Data collapse of pressure against density.  (a) standard
DPD model, for $\rho=1$--10 and $A=0$--50.  Straight line is fit to
data given by Eq.~(\ref{eqpstd}) in the text.  (b) Many-body DPD
model, for $\rho=1$--10, $A=0$--50, $B\rd^4=0$--10 and $\rd=0.5$--1.0.
The ordinate is the function $F = (p-\rho-\alpha A\rho^2-2\alpha
B\rd^4\rho^3)/B\rd^4$.  Plotted this way, the data collapse onto
quasi-universal straight lines, where the slope depends primarily on
$\rd$.  This can be expressed as a quasi-universal EOS in
Eq.~(\ref{eqpmany}) in the text.\label{fig1}}
\end{figure}

\section{A specific model}
I now describe in more detail a specific application of these ideas to
set up a DPD model which exhibits vapour-liquid coexistence.  Before
this though, there is one more technical point to discuss.  

To stabilise the vapour-liquid interface, it is not sufficient just to
have a van der Waals loop in the EOS; one must also give consideration
to the ranges of the interactions.  Thus simple many-body DPD with a
single range may not have a stable interface as discussed by
Pagonabarraga and Frenkel \cite{PF}.  The trick employed here is to
take the standard DPD model, make the soft pair potential attractive,
and add on a repulsive many-body contribution with a \emph{different}
range $\rd<\rc$.  Furthermore I choose the simplest form of the
many-body repulsion, namely a self energy per particle which is
quadratic in the local density.

In terms of force laws, I take the standard DPD model as
specified in Eq.~\eqref{feq} with $A<0$, and add a many-body force
law of the form 
\begin{equation}
\FijC = B(\rhobar_i + \rhobar_j)\,\wC(r_{ij})\,\eij
\label{mb2feq}
\end{equation}
where $B>0$.  This is Eq.~\eqref{mbfeq} with $A(\rhobar)=2B\rhobar$.
The weight function in this is chosen to be $\wC(r)=(1-r/\rd)$ for
$r<\rd$. This means that $\wrho(r)=15/(2\pi\rd^3)(1-r/\rd)^2$
(normalised for three dimensions) is used to construct the local
density, and $\alphaMF = \pi\rd^4/30$ for this particular interaction.

In terms of potentials, this model can be interpreted as follows.
Define a generalised weight function of range $R$ via
\begin{equation}
\wrho(r;R) = {15}/{(2\pi R^3)} (1-{r}/{R})^2.
\end{equation}
Then define \emph{two} local densities $\rhobar$ and $\rhobarbar$,
constructed using this weight function with $R=\rc$ and $R=\rd$
respectively.  The self energy per particle for this specific model
can be written as
\begin{equation}
u = (\pi/30)A\rhobar + (\pi\rd^4/30) B\rhobarbar{}^2.
\end{equation}
This is at most quadratic in the local densities, and thus the model
could be written out explicitly in terms of two- and three-body
interaction potentials.  From this, the mean field EOS is
\begin{equation}
\pMF = \rho\kT + (\pi/30)(A + 2B\rd^4\rho)\rho^2.
\label{pmfeq}
\end{equation}
Thus, with $A<0$ and $B>0$, this EOS has the potential to contain a
van der Waals loop.  The actual EOS differs from this systematically
as I now describe.

\begin{table}
\begin{tabular}{c|ccccc}
$\rd$&0.50&0.65&0.75&0.85&1.00\\
\hline
$c$&4.0(5)&4.1(1)&3.07(5)&2.08(5)&1.29(5)\\
\end{tabular}
\caption[?]{Density correction `constant' for many-body term in
measured EOS, as a function of $\rd$.  There is little significant
dependence on other parameters $A$, $B$ and $\rho$ (see eg
Fig.~\ref{fig1}(b)).  Figures in brackets are estimates of the error
in the final digit.\label{tab1}}
\end{table}

\begin{figure}
\begin{center}
(a)\includegraphics{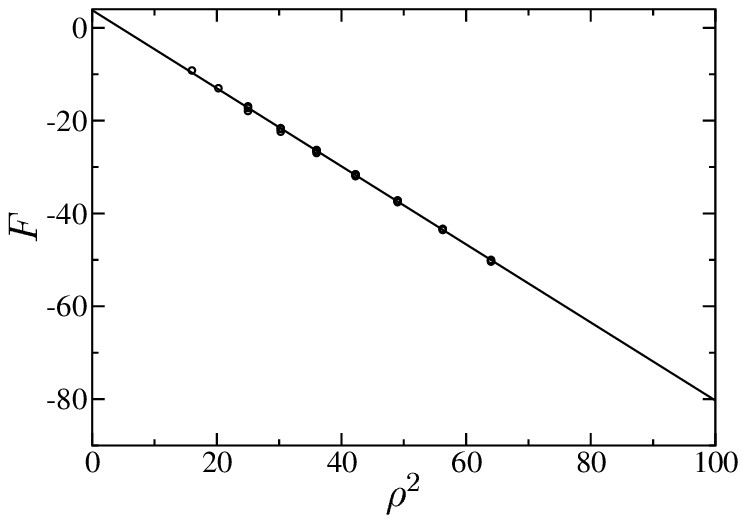}\\
(b)\includegraphics{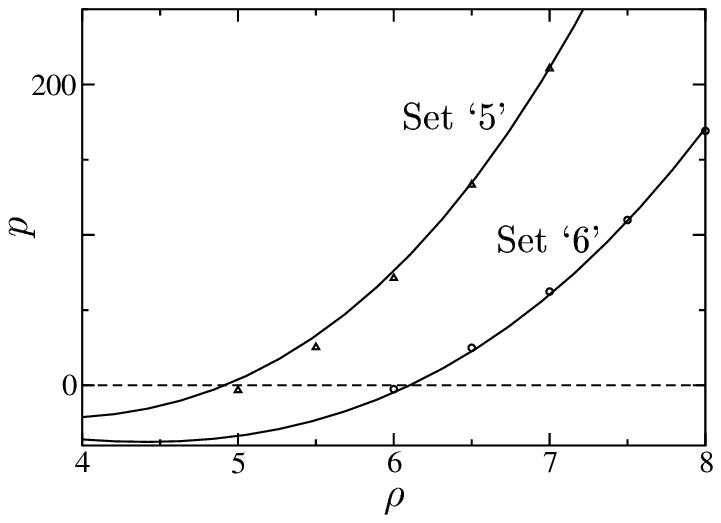}
\end{center}
\caption[?]{(a) Data collapse of pressure against density, for
$\rhoL<\rho\alt8$, $A<0$ and $|A|=20$--40, $B=25$ or 40, and
$\rd=0.75$.  The ordinate is the function $F = (p-\rho-\alpha
A\rho^2-2\alpha B\rd^4\rho^3)/B\rd^4$ Straight line is fit to data
given by Eq.~(\ref{eqpsel}) in the text.  (b) Pressure as a function
of density for the two selected parameter sets in Table~\ref{tab2}.
The lines are the predictions of the fitted EOS, Eq.~(\ref{eqpsel}) in
the text.\label{fig2}}
\end{figure}

\section{Simulation results}
I now explore by simulation the properties of the above model.  First
I examine the actual EOS, then vapour-liquid coexistence and the
properties of a stable vapour-liquid interface, and finally illustrate the
potential application of the method with a simple pendant droplet
simulation.  Typical simulations presented here are in simulation
boxes of size $10^3$ (units of $\rc$).

\subsection{Equation of state}
For $B=0$, the standard DPD model, the simulations recover the
accepted equation of state Eq.~\eqref{eqpstd} with very small corrections
of $\sim\rho^3$.  Results are shown in Fig.~\ref{fig1}(a).

For $A>0$ and $B>0$ a large number of simulations were performed.
After some experimentation, the data was found to collapse to the
following quasi-universal behaviour,
\begin{equation}
p=\rho+\alpha A\rho^2+2\alpha B\rd^4\rho^2(\rho-c).\label{eqpmany}
\end{equation}
where $\alpha$ takes the same value as for standard DPD (and thus this
expression contains the correct $B=0$ limit), and $c$ is an empirical
correction to the density that appears in the many-body term.  This
should be compared to the mean field prediction in Eq.~\eqref{pmfeq}.
I find that $c$ depends predominantly on $\rd$ according to
Table~\ref{tab1}.  A representative sample of the data is shown in
Fig.~\ref{fig1}(b).

For some parameter sets, the temperature was found to show strong
deviations from the nominal $\kT=1$, as a result of instabilities in
the integration algorithm.  Results were only kept if the measured
$\kT$ lay within 10\% of the nominal value.  These problems occur if
the repulsion amplitudes are too large, or the densities too high, or
$\rd$ too small.  The integration algorithm is that described in Groot
and Warren \cite{GW}, with a time step $\Delta t = 0.05$ and $\lambda
= 1/2$.  The instabilities can be vanquished by making $\Delta t$
smaller.

The measured equation of state is therefore quite close to the
predicted mean field equation of state.  The main difference is a
correction to the density dependence of the many-body term.  This is
expected since the pair correlation function $g(r)<1$ where the
repulsions are strongest, and thus $\average{\rhobar_i}<\rho$.  This
effect has also been checked in simulations by monitoring the mean
value of the local density, with results similar to those
reported by Trofimov \etal\ \cite{TNM}.

\begin{figure}
\begin{center}
\includegraphics{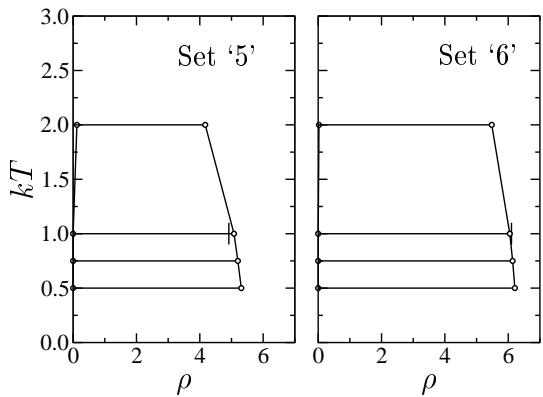}
\end{center}
\caption[?]{Density-temperature phase diagrams for the two parameter
sets in Table~\ref{tab2}.  Shown horizontal are tielines computed from
vapour-liquid interface profiles, at several temperatures (keeping $A$
and $B$ fixed).  The short vertical line on the $\kT=1$ tieline is the
point where the pressure vanishes according to the fitted EOS,
Eq.~(\ref{eqpsel}) in the text.\label{fig3}}
\end{figure}

\subsection{Vapour-liquid coexistence}
For vapour-liquid coexistence I set $A<0$ and $B>0$ so that there is a
van der Waals loop in the EOS.  Phase separation is found in a range
of densities $\rhoV<\rho<\rhoL$ where $\rhoV$ and $\rhoL$ are the
vapour and liquid coexistence densities.

In principle, integration of the EOS gives the free energy density
from which predictions can be made about $\rhoV$ and $\rhoL$.
Unfortunately, the EOS must deviate from the above fitted form for
$\rho\ll 1$, therefore the vapour phase is inadequately characterised.
For applications, one is most interested in $\rhoL\agt1$ in
coexistence with a very dilute vapour.  If this is true it is much
easier to use the EOS to predict the point where the pressure vanishes
as an estimate of the coexisting liquid phase density, thus $p(\rhoL)
= 0$.  Using this, one expects liquid densities in the range
$\rhoL\sim5$ for $-A\sim B\sim30$.  From here on I have set the
range of the many-body repulsion to $\rd=0.75$ as a mid-range value
determined above.

Since the above EOS was measured for $A>0$, one has to be careful to
check that the scaling collapse still holds.  One cannot easily
measure the EOS within the phase separation region, since it is hard
to maintain a stable uniform density.  Therefore the EOS has been
characterised for $\rho>\rhoL$.  A similar data collapse is found to
the previous section, as is shown in Fig.~\ref{fig2}(a).  In this
case, the EOS can be fitted by
\begin{equation}
p=\rho+\alpha A\rho^2+2\alpha B\rd^4(\rho^3-c\rho^2+d)\label{eqpsel}
\end{equation}
where $\alpha=0.101(1)$ as before, $c=4.16(2)$ and $d=18(1)$.  The
value of $c$ is similar to the value obtained previously ($c=3.07(5)$,
Table~\ref{tab1}).  There is an additional offset term $d$ which is
about 10\% of the density correction term $c\rho^2$ in the region of
interest.  

\begin{figure}
\begin{center}
\includegraphics{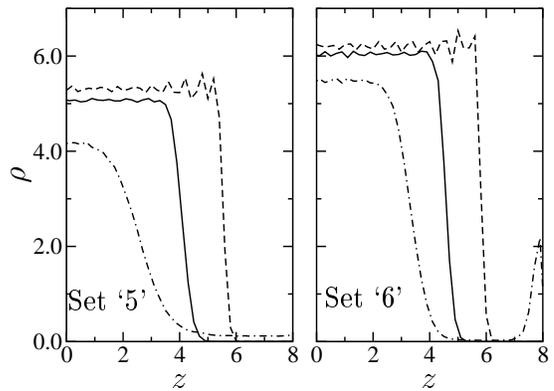}
\end{center}
\caption[?]{Interfacial density profiles for the two parameter sets in
Table~\ref{tab2}, for several values of the temperature (keeping $A$ and $B$
fixed): $\kT=1$ (solid line), $\kT=2$ (chained line), $\kT=0.5$
(dashed line).\label{fig4}}
\end{figure}

\begin{table}
\begin{tabular}{c|ccc|cccc}
set&$A$&$B$&$\rd$&$\rhoL$&$w$&$\sigma$&$\partial p/\partial\rho$\\
\hline
`5'&$-40$&40&0.75&5.08(1)&0.78(5)&4.95(3)&49(2)\\
`6'&$-40$&25&0.75&6.05(1)&0.66(3)&7.45(4)&47(2)\\
\end{tabular}
\caption[?]{The two parameter sets used in subsequent simulations.
The sets are distinguished by the different values of the
liquid densities $\rhoL$.  The coexisting vapour density
$\rhoV\ll1$, so these parameters are suitable for free surface
simulations.  Also shown are the interface width $w$, surface tension
$\sigma$, and compressibility at $\rho=\rhoL$ estimated from the EOS.
All results are at $\rc=\kT=1$.\label{tab2}}
\end{table}

Although a wider parameter space was explored, I concentrate here on
two parameter sets that were selected for more detailed work.  These
parameter sets are given in Table~\ref{tab2} (first three columns).
Fig.~\ref{fig2}(b) shows the prediction of the EOS,
Eq.~(\ref{eqpsel}), compared directly against the measured pressures,
for these two parameter sets.

For these two parameter sets, the coexisting vapour and liquid
densities were determined from the vapour-liquid interface profile
simulations described in the next section, and are shown as a function
of $\kT$ in Fig.~\ref{fig3}.  Where $\kT\ne1$ in these simulations,
the values of $A$ and $B$ are left at the values in Table~\ref{tab2},
in other words $A$ and $B$ are regarded as absolute interaction
energies.  Also shown in Fig.~\ref{fig3} are the appropriate solutions
of $p=0$ using the EOS in Eq.~(\ref{eqpsel}).

It is clear that the difference $\rhoL-\rhoV$ gets smaller as
$T$ increases, as one approaches the expected vapour-liquid critical
point. At $\kT\alt1$, $\rhoV\ll1$ indicating the vapour phase is
virtually devoid of particles. At $\kT=1$ the solution to $p=0$ for
the EOS gives a good estimate of the density of the fluid phase.

I have also used the EOS to estimate the compressibility $\partial
p/\partial\rho$ at $\rho=\rhoL$, and the values are shown in
Table~\ref{tab2}.  Although the precise value is not important, the
fact that $\partial p/\partial\rho\gg1$ at the coexisting fluid
density (where $p\approx0$) shows that the fluid phase is
relatively incompressible, similar to a real liquid.

\begin{figure}
\begin{center}
\includegraphics{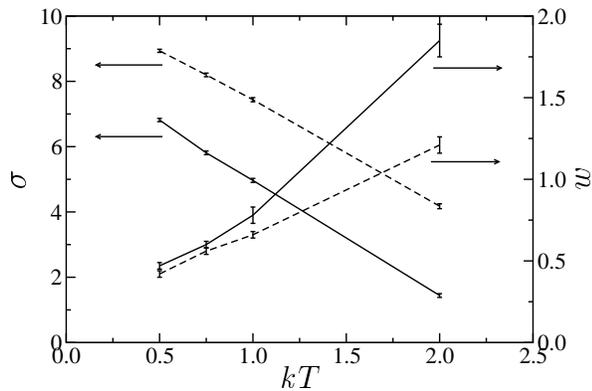}
\end{center}
\caption[?]{Interfacial tension $\sigma$ (left hand axis) and
interface width $w$ (right hand axis), as a function of temperature
$\kT$, for the two parameter sets in Table~\ref{tab2} (keeping $A$ and $B$
fixed): set `5' (solid lines) and set `6' (dashed lines).\label{fig5}}
\end{figure}

\subsection{Vapour-liquid interface}
Simulations of the vapour-liquid interface were undertaken, by taking
an equilibrated volume of fluid in a periodic box, at a density close
to the $p=0$ limit, and removing the particles in one half of the box.
The system was allowed to evolve until an equilibrium density profile
was obtained.  Interface profiles and surface tension values were
measured as described for fluid-fluid interfaces \cite{CLRW}.  For
measurement of density profiles, it was necessary to stop the
interface drifting over time.  This was achieved by inserting a thin
slab of `frozen' particles at one end of the box.

Fig.~\ref{fig4} shows the interface profiles obtained this way for the two
selected parameter sets in Table~\ref{tab2}.  These are shown at several
different values of $\kT$ keeping $A$ and $B$ fixed.  The limiting
densities on either side of the interface were used to construct the
tielines discussed in the previous section (Fig.~\ref{fig3}).  As the
temperature is increased, the interfacial width $w$ gets broader, and
the surface tension $\sigma$ drops.  Results for $\sigma$ and $w$ are
shown in Fig.~\ref{fig5}.  The width $w$ was quantified by calculating the
maximum slope, and normalising to the coexistence densities, thus
\begin{equation}
w=\frac{\rhoL-\rhoV}{\mathrm{max}\,|d\rho/dz|}.
\end{equation}
The surface tension is determined from the standard mechanical
definition of the pressure tensor \cite{ATbook}.  Note again
that there is no problem with the many-body origin of the force laws.
The actual forces enter the calculation in exactly the same way as
standard DPD.

Low temperature favours a sharp interface, but if the temperature is
too low, oscillations develop in the profile on the liquid side of the
interface.  This can be seen most clearly in Fig.~\ref{fig4} for
$\kT=0.5$.  The system has crossed a Fisher-Widom line in the phase
diagram, and a freezing transition is almost certainly nearby.  The
relative amplitude of the oscillations can be measured, and they are
typically 10\% of the bulk density at $\kT=0.5$, but $<2$\% for
$\kT\ge0.75$, at least for the two sets of parameters studied here.

Thus I conclude that the two parameter sets given in Table~\ref{tab2}
provide for a sharp vapour-liquid interface, at $\kT=1$, with
virtually no particles in the vapour phase.  They are thus well suited
to model free surfaces.  Table~\ref{tab2} also contains the
measured interfacial properties.

\begin{figure}
\begin{center}
\includegraphics{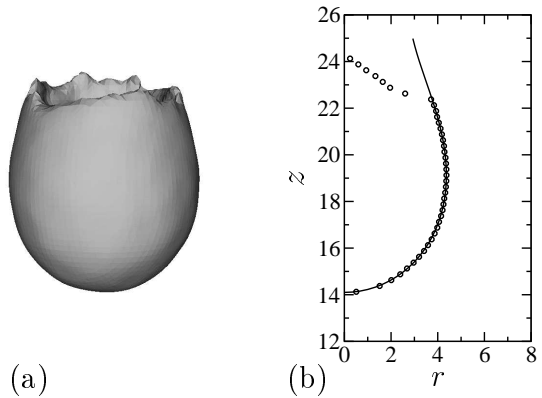}
\end{center}
\caption[?]{Pendant droplet problem: (a) isosurface cut through $3d$
density field at $\rho=\rhoL/2$, showing the drop profile, and (b)
drop radius $r$ as a function of height $z$, computed by the method
described in the text.  The solid line is the Young-Laplace equation
prediction, the circles are the measured profile.  The `frozen'
support particles at the top of the drop give a ragged edge to the top
of the isosurface in (a), and distort the measured profile for
$z\agt23$ in (b).  Parameters correspond to set `6h' in
Table~\ref{tab3}.\label{fig6}}
\end{figure}

\begin{table}
\begin{tabular}{c|cc|ccc|cc}
set&$g$&$\rhoL$&$\Delta E$&$\beta$&$1/H(\beta)$&%
$\sigma_{\mathrm{shape}}$&$\sigma_{\mathrm{exact}}$\\
\hline
`5a'&0.022&5.08&8.80(5)&0.378&0.57(3)&4.9(3)&4.95(3)\\
`6f'&0.025&6.05&9.00(5)&0.378&0.57(3)&7.0(5)&7.45(4)\\
`6h'&0.033&6.05&8.75(5)&0.404&0.52(2)&7.9(4)&7.45(4)\\
`6g'&0.030&6.05&8.70(5)&0.431&0.48(4)&6.6(5)&7.45(4)\\
\end{tabular}
\caption[?]{Pendant drop profile data.  Interaction parameters are
taken from Table~\ref{tab2}, according to the first digit of `set'.
The penultimate column is the surface tension computed from the drop
shape, and the final column is the `exact' surface tension from
Table~\ref{tab2}, computed by integration of the pressure tensor
through a planar interface.\label{tab3}}
\end{table}

\subsection{Pendant droplet simulation}
As an example application, I look at the classic pendant drop problem.
The procedure is very similar to the one adopted for the DPD
multiphase fluid model \cite{CLRW}.

To set up the pendant droplet, a volume of fluid at a density close to
the equilibrium liquid density was equilibrated, then replicated to
construct a cylindrical column with the axis parallel to the
$z$-direction.  A `support' was constructed by `freezing' particles in
a thin slice at the top of the column.  A gravitational body force $g$
was included by adding a constant force per particle directed along
the $z$-direction away from the support.  When the system reaches
equilibrium, the liquid forms a pendant droplet suspended from the
support particles.  In equilibrium, the drop profile (radius as a
function of height) was obtained as described below.  The whole
simulation takes a couple of minutes on a modern workstation.  The
droplet contains typically $\sim3000$ particles.

The profile was determined as follows.  A $3d$ mesh was introduced
with a resolution typically $\le0.5\rc$ (higher resolution was
employed in the $z$-direction). The local particle density in each
mesh volume element was computed by averaging over a period of time.
This gives a $3d$ density field.  The droplet can then be imaged as an
isosurface or level cut through this density field, and a typical
result is shown in Fig.~\ref{fig6}(a).

To determine the drop radius as a function of height, the density
field was divided (or `segmented') into occupied and unoccupied cells
according to whether $\rho(\rvec)>\rhoL/2$ or not.  The number of
occupied cells at each height $z$ was used to compute the
cross-sectional area of the droplet at that height, and therefore the
drop radius as a function of $z$.  A typical drop profile is shown in
Fig.~\ref{fig6}(b).

This indirect procedure to determine the droplet radius eliminates two
possible artefacts.  Firstly it removes the blurring of the base of
the drop by the interface profile, which would otherwise be
$\sim0.7$--$0.8\rc$.  Secondly it eliminates effects due to the
variation of fluid density with height which might otherwise introduce
a systematic error, if the mean particle number density as a function
of height was computed directly.  Such a variation of density with
height is to be expected, since the fluid responds to the varying
pressure field through the EOS (ie it is still a compressible fluid,
even if only weakly so).

The drop profile was analysed by normalising with respect to the
maximum diameter $\Delta E$, and comparing with a set of precalculated
profiles as the Bond number $\beta = \rhoL g b^2/\sigma$ varies
(where $b$ is the radius of curvature of the base of the droplet).
The profiles are calculated from the Young-Laplace equation as described in
earlier work \cite{CLRW}.  From the best-fit $\beta$ value, the
surface tension can be computed from $\sigma=\rhoL g \Delta E^2/H$
where $H(\beta)$ is a dimensionless function computed numerically.

Table~\ref{tab3} shows the quantities computed for several drops, for
both parameter sets, and for several values of $g$.  Although the
surface tensions determined this way are not very precise, they are
all consistent with the accurate values calculated directly from the
interfacial profiles.  The drop profiles all match the measured
profiles quite accurately, see for example Fig.~\ref{fig6}(b).

\section{Discussion}
The model developed here can be discussed in several contexts.
Firstly, it is a new simulation method for fluid mechanics problems
involving liquids with free surfaces.  For example, the above pendant
droplet problem is a test of the static force balance and the results
show that the DPD fluid obeys the Young-Laplace equation in a non-trivial
geometry.  One can conclude that this particular version of many-body
DPD offers a viable route for solving capillary problems such as the
distribution of liquids in porous materials.  It is clearly possible
to address dynamic force balance situations too, but these will
require further testing and parametrisation, particularly for the
notorious problem of contact line dynamics.

Secondly, now that vapour-liquid equilibrium is achieved for a basic
soft sphere model, one can `dress' the liquid up in various ways such
as making the liquid particles into polymers or model amphiphiles.  In
this way, new methods can be constructed to simulate complex fluids
with an \emph{implicit} solvent.  These developments are the subject
of ongoing investigations and will be reported separately.

In a third context though, the re-interpretation of many-body DPD as a
fluid whose potential energy is built out of local-density dependent
one-particle self energies is quite novel from the point of view of
liquid state theory.  Most previous work has concentrated on fixed
pair potentials with hard cores, and only minor attention has been
paid to soft potentials or density-dependent pair potentials.  The
present work though goes some way beyond these existing ideas.

It has long been recognised that an arbitrary
$U(\lbrace\rvec_i\rbrace)$ can be expanded as a sum over
density-independent one-body, two-body (pair potential), etc, terms.
Normally the one-body terms, or self energies, are harmless constants
which can be discarded, and most of the phenomena observed for liquids
can be captured by truncating the expansion at the pair potential
level.  If one allows the pieces in such an expansion to acquire a
density dependence though, then the one-body self energy is no longer
necessarily a constant, and it is no longer necessary to go to the pair
potential level to see interesting physics.  Many-body DPD as
described here is an example of precisely this.

The phase behaviour of the present model is also potentially very
interesting.  By analogy with related soft-core systems such as the
Gaussian core model \cite{gcm,LBH}, and models for polymers of various
architectures \cite{stars,LHLL}, the particles in the original DPD
model are expected to freeze into a variety of ordered phases at low
temperatures and intermediate densities, with a re-entrant fluid phase
at high densities.  The version of many-body DPD presented in this
paper is constructed to have a significant vapour-liquid coexistence
region, as shown in Fig.~\ref{fig3}, but the low temperature ordered
phases are presumably still present, as indicated by the presence of
oscillations in the liquid side of the vapour-liquid interface in
Fig.~\ref{fig4}.  In such a case, the collision between the
vapour-liquid transition and these ordered phases could prove to
generate rather unusual phase behaviour, and the low temperature
properties of many-body DPD models may well be worth further
examination.

I thank R. D. Groot for many discussions in the early stages of
development of the many-body DPD model.

\bibliography{mbdpd}

\end{document}